\documentclass[12pt,preprint]{aastex}
\usepackage{emulateapj5}
\def\gtrsim{\mathrel{\hbox{\rlap{\hbox{\lower4pt\hbox{$\sim$}}}\hbox{$>$}}}}
\let\ga=\gtrsim
\def\lesssim{\mathrel{\hbox{\rlap{\hbox{\lower4pt\hbox{$\sim$}}}\hbox{$<$}}}}
\let\la=\lesssim
                                \begin{document}

                                \title{
The Merger in Abell 576: A Line of Sight Bullet Cluster? 
                                }

                                \author{
 Renato A. Dupke, Nestor Mirabal, Joel N. Bregman \& August E. Evrard
                                }

                                 \affil{
University of Michigan, Ann Arbor, MI 48109-1090
                                }

                                \begin{abstract}
 
Using a combination of {\sl Chandra} and {\sl XMM} observations, we  confirmed the presence of a significant velocity gradient 
along the NE/E--W/SW direction in the intracluster gas of the cluster Abell 576. The results are consistent 
with a previous {\sl ASCA} SIS analysis of this cluster. 
 The error weighted average over ACIS-S3, EPIC MOS 1 \& 2 spectrometers for the maximum velocity
 difference  is $>$3.3$\times$10$^3$ km~s$^{-1}$ at the 90\% confidence level, 
 similar to the velocity limits estimated indirectly for the ``bullet'' cluster (1E0657-56). 
The probability that the velocity gradient is generated by standard 
random gain fluctuations with {\sl Chandra} and {\sl XMM} 
is $<$0.1\%. The regions of maximum velocity gradient are in CCD zones that have the lowest temporal
gain variations. It is unlikely that the velocity gradient is due to Hubble distance 
differences between projected clusters (probability$\la$0.01\%).
We mapped the distribution of elemental abundance ratios  
across the cluster and detected a strong chemical discontinuity using the abundance ratio 
of silicon to iron, equivalent to a variation from 100\% SN Ia iron mass fraction 
in the West--Northwest regions to 32\% in the Eastern region.  The ``center'' of the 
cluster is located at 
the chemical discontinuity boundary, which is inconsistent with the 
radially symmetric chemical gradient 
found in some regular clusters, but consistent with a cluster merging scenario.
We predict that the velocity gradient as measured will produce a variation of the CMB temperature towards
the East of the core of the cluster that will be detectable by current and near-future bolometers. 
The measured velocity gradient opens for the possibility that this cluster is passing through 
a near line-of-sight merger stage where the cores have recently crossed.

                                \end{abstract}

                                \section{
Introduction
                                }
The characterization of the internal dynamics of the intracluster medium is very important for determining 
 the evolutionary stage of galaxy clusters (Beers et al. 1982), to study cluster formation and  
to assess the systematics of using clusters of galaxies as cosmological tools. 
The presence of surface brightness features detected by the {\sl Chandra} satellite such as 
cold fronts, shock fronts and X-ray cavities shows that the intracluster gas (ICM) 
is often dynamically
active. Furthermore, departure from assumptions such as hydrostatic equilibrium has 
been justified theoretically 
(e.g. Kay et al. 2004; Rasia, Tormen \& Moscardini 2004, 2006; Pawl, Evrard \& Dupke 2005), 
but detection of bulk gas velocities became possible only with the launch of the {\sl ASCA} satellite and more 
recently with the spectrometers on-board {\sl Chandra} and {\sl XMM-NEWTON}.

The key ingredient to quantify the level of activity is the determination of gas 
bulk (or turbulent) velocities. In order to assess the gas dynamics we would ideally like 
to have a ``direct'' measurement of intracluster gas 
velocities. Since the intracluster medium is enriched with heavy elements, this can be done, 
for example, by measuring the Doppler shift of the spectral lines in X-ray frequencies 
(Dupke \& Bregman 2001a,b)
or by measuring changes in line broadening due to turbulence 
(Inogamov \& Sunyaev 2003; Sunyaev, Norman \& Bryan 2003; 
Pawl et al. 2005). The former can currently be done only if there are enough photon counts 
within the spectral lines, if the instrumental gain is stable and well known and if 
the instrument has good spectral 
resolution. Doppler shift analysis of clusters started with the {\sl ASCA} satellite, which set 
constraints on bulk velocity 
gradients in 14 nearby clusters (Dupke \& Bregman 2001a,b, 2005). However, {\sl ASCA}
relatively high gain temporal variation limited velocity constraints to $\ge$2000 km/s, so that it is 
crucial to corroborate and improve previous measurements of velocity gradients found 
in the {\sl ASCA} sample with other instruments if we wish to investigate intracluster gas dynamics.

The higher stability and better spectral resolution of ACIS-S3 and MOS 1 \& 2 on-board {\sl Chandra}
and {\sl XMM-Newton} satellites provide, currently, a unique opportunity to improve the constraints on ICM velocity
gradients, allowing a factor of $\ga$ 2 improvement in the uncertainties of velocity measurements.
The two clusters found to have the most significant velocity gradients with {\sl ASCA} were 
the Centaurus cluster (Abell 3526) and Abell 576. Velocity gradients have been confirmed in the Centaurus 
cluster in two off-center {\sl Chandra} pointings (Dupke \& Bregman 2006, hereafter DB06; however, see Ota et al. 2007) 
and here we show a combined 
velocity analysis of {\sl Chandra} and {\sl XMM-Newton} pointings of Abell 576.

Abell 576 is a richness class 1 cluster with relatively low (T$\sim$4 keV) central 
gas temperatures and average metal 
abundances (e.g. Rothenflug et al. 1984; David et al. 1993; Mohr et al. 1996). It has an optical redshift of 0.0389. ASCA velocity analysis of this cluster found a 
significant velocity gradient ($>$4000 km/s, Dupke \& Bregman 2005 (hereafter DB05)). 
Evidence for dynamic activity in this 
cluster has been put forward in previous analyses. Rines et al. (2000) determined the mass profile of 
A576 using the infall pattern in velocity 
space for more than 1000 galaxies in a radius of 4 h$^{-1}$ Mpc from the cluster's center. They found that 
the mass of the central Mpc was more than twice of that found from X-ray measurements, suggesting that
nonthermal pressure support may be biasing the X-ray derived mass. 
Additional evidence for strong departures from hydrostatic equilibrium comes
from energy excess of the X-ray emitting gas with respect to the galaxies (Benatov et al. 2006).
These characteristics can be partially explained by non-thermal pressure 
support and significant departures from spherical symmetry due to a line of sight merger. 
Mohr et al. (1996), using
galaxy photometric data, found a high velocity tail separated by $\sim$3000 km/s from the cluster's mean. 

Kempner \& David (2004), hereafter KD04,
analyzed a Chandra observation of the core of this cluster and found brightness edges corresponding 
to mild jumps in gas density and pressure roughly in the N-S direction. The X-ray image
of the cluster also shows an ``arm'' extending 
to the SW and mild evidence of wakes (``fingers'') in the N-NW direction (Figure 1a).
The authors suggested 
that the core substructures are caused by a current merger with 
core velocities of $\sim$ 750 km~s$^{-1}$, to maintain the gas confined across the surface brightness 
edge towards the N. In their scenario the merging cluster came in from 
the direction of the ``fingers'' (N-NW), has passed the core of the main cluster, created the SW and W edges 
and is now near the second core passage. In this paper, we perform a velocity analysis 
of Abell 576 using the full field of view covered by 
{\sl Chandra}'s ACIS-S3 and combine it with two
{\sl XMM}'s EPIC MOS 1 \& 2 from two observations, specifically tailored to minimize random 
gain variations across the CCDs. We also present an analysis of the distributions 
of intracluster gas temperature, 
velocity and individual elemental abundances and use them to determine the evolutionary stage of this cluster. 
 All distances shown in this work
are calculated assuming a H$_0=70$ km~s$^{-1}$Mpc$^{-1}$ and 
$\Omega_0=1$ unless stated otherwise. 

\section{Data Reduction and Analysis}
\subsection{Chandra}

Abell 576 was observed for 39 ksec on Oct 2002 centered on ACIS-S3. Nearly a fourth of the observation was affected by
flares and we here show the analysis of the unaffected initial 29 ksec of observation. We used CIAO 3.2.1 with 
CALDB 3.1.0 to screen the data.
The data were cleaned using standard procedure\footnote{http://cxc.harvard.edu/ciao/guides/acis\_data.html}.
Grades 0,2,3,4,6 were used. ACIS particle background was cleaned as prescribed for VFAINT mode. 
A gain map correction was applied together with PHA and pixel randomization. Point 
sources were extracted and the background used
in spectral fits was generated from blank-sky observations using
the {\tt acis\_bkgrnd\_lookup} script. 

In order to obtain a overall distribution of the spectral parameters we developed an 
``adaptive smoothing'' code that selects regions for spectral extraction based on a pre-determined 
minimum number of counts, which 
for the cases shown here was 5000 cnt/cell. The overlap of extraction regions is therefore stronger in 
the low surface brightness regions, away from the cluster's core. We also excluded the CCD 
borders by $\sim$ 1$^\prime$ to avoid ``border effects'', characteristic of these type of
codes.
The responses were created for each individual region with the 
CIAO tools {\it makeacisrmf} and {\it mkwarf}. 
Spectra and background spectra were generated and fitted with XSPEC V11.3.1 (Arnaud 1996) with an absorbed 
{\tt VAPEC} thermal emission models. 
Metal abundances are measured relative to the solar photospheric values of 
Anders \& Grevesse (1989).
Galactic photoelectric absorption was incorporated using the WABS 
model (Morrison \& McCammon  1983). Redshifts were determined through spectral fittings using 
a broad energy range. 
In the spectral fits we fixed the Hydrogen column density N$_H$ at its corresponding Galactic value
of 5.7$\times$10$^{20}$ cm$^{-2}$.
Spectral channels were grouped to have at least 20 counts/channel. Energy 
ranges were restricted to 0.5--9.0 keV. The spectral fitting parameter errors listed here 
are 1-$\sigma$ unless stated otherwise. For all spectral fittings used here we applied the recursive process 
to find the best-fit redshift with 
"true" $\chi^{2}$ minimum as described in DB05.

\subsection{{\sl XMM}}
Abell 576 was observed with {\sl XMM--Newton}  on 2004 March 23 for $\sim$ 22 ksec. A second observation
was obtained a few days later on 2004 March 27 for a total of $\sim$ 20 ksec. The observations were 
planned in such a way as to overlap the cluster's core, while providing
sufficient coverage on the northeast and southwest of
the cluster, which were the regions expected to have the strongest velocity gradient from a previous 
{\sl ASCA} observation (DB05) (Figure 1b).  This observational strategy was designed to minimize the impact 
that spatial variations of the gain (conversion between pulse height and energy of an incoming 
photon) has on redshift measurements.

Initial inspection of the EPIC MOS and PN data
revealed a number of strong background flares. 
In order to exclude these periods of high
background, good time intervals were produced from events where the threshold did not
deviate more than 3 $\sigma$ from the extrapolated mean count rate in the 10--15 keV band. In addition,
only events satisfying grade patterns $\leq$12 have been used.
The effective exposure times after removal of background flares
correspond to $\sim$ 12 ksec (55\% of the total) for the first pointing, and $\sim$ 16 ksec (80\% of the total)
for the second. Using these cleaned event lists,
background spectra were produced from several source-free regions on the detector away from the source. 
Blank-sky backgrounds were also used for comparison with no significant changes in the resulting best-fit 
parameters.
The data presented here were processed
with {\em XMM-Newton} Science Analysis System  SAS 6.0.0. Response files for each region
have been generated using the SAS tasks {\it rmfgen} and {\it arfgen}. Bright point sources were 
extracted and the spectral fitting routine was identical to that used 
with the {\sl Chandra} data described in the previous section. Only MOSs 1 \& 2 were used
because of the high number of interchip boundaries within our regions of 
interest in the PNs, which would affect significantly the estimation of gain fluctuations. Furthermore, 
the loss of data due to flares was especially 
strong for the PNs. Despite the relatively small number of counts the {\sl XMM} observation helped 
to constrain the spectral parameters derived from {\sl Chandra}.

 \section{
 Projected Temperature and Velocity Contour Maps
                                 }
The resulting temperature and velocity distributions from the adaptive smoothing routine applied to the 
{\sl Chandra} data are shown in Figures 2a,b. The colors are chosen in a way as to show 
the average 1-$\sigma$ variations. 

The temperature map shows that the cluster's core regions is relatively cold ($\sim$ 3.5 keV)
and has an overall asymmetric distribution. 
The coldest region ($\sim$3.0 keV) is not found in the core but at the NE region.
Interestingly, it can also be seen that the highest gas temperature is found 2$^\prime$--3$^\prime$ 
towards the NW 
direction and reaches 
$\approx$5 keV. This was not noted in KD04, due to their choice of orientation 
for selection of the extraction regions.
Overall, the temperature distribution follows roughly a configuration where a cold core is surrounded by a 
hotter elliptical ring  elongated along the NW-SE direction (shown by the dashed lines in Figure 2a). 
There are also marginal indications that the temperature
decreases again at regions $>$3$^\prime$ to the E and S directions.

The velocity map (Figure 2b) is not smooth and shows higher velocities 
in the Southern regions, and a clear zone of lower 
redshifts to the NE that extends to the central region. 
Even though the highest redshift zone is apparently in the SE corner,  
analysis of the error map in Figure 2c shows that region has very high 
uncertainties. 
To find the regions of maximum significance of velocity measurements, in each cell we divided the 
difference of the best fit redshift from the average over the CCD (denoted by $<>$) by the error
of the measured redshift $\delta$z.
i.e., $\frac{z-<z>}{\delta z}$ (see DB05 \& DB06 for details).
We denote this error-weighted-deviation simply as deviation significance and plot its
color contours in Figure 2c. In 
Figure 2c the black and white represent negative and positive velocities, respectively, with respect to the 
CCD average velocity. 
The magnitude of the deviation significance shows how significant the velocity structure is. 
We can see that the region of maximum negative significance is located slightly to the E of the cluster center.
There is also a region of marginally higher positive significance ($\sim$3$\sigma$) to the SW, in good agreement with 
previous observations with {\sl ASCA}. Based on these two deviation significance peaks we selected two regions 
for a more detailed study, shown in
Figure 1b as black rectangles; a high (redshifted) and low (blueshifted) redshift regions, hereafter called 
{\b SOUTH} and {\b EAST}, respectively. Although the cluster core seems to be included in the 
blueshifted zone in both
{\sl Chandra}, {\sl XMM} (and was also in {\sl ASCA} SISs) we, conservatively, avoid including it in our velocity analysis due to 
modeling uncertainties (see DB05 for a more extended discussion on the effects of multiple models 
in the best-fit redshift with the technique used here). Below we explore in more detail the
spectral analysis of these regions. 

 \section{
 Chandra and XMM Velocity Analysis of Selected Regions  
}

The best-fit gas temperatures, iron abundances and velocities for the two regions with highest deviations from the average 
redshift are 
plotted in Figure 3a and listed in Table 1. The spectra corresponding two these two regions are shown in 
Figures 3b,c,d for different spectrometers.
Individual spectral fits of these regions show very similar gas temperatures, with an error weighted average 
of 3.87$\pm 0.11$ keV for {\b SOUTH} and 4.00$\pm 0.11$ keV for {\b EAST}, and also similar iron 
abundances, with an error weighted average 
of 0.54$\pm 0.06$ solar for {\b SOUTH} and 0.52$\pm 0.05$ solar for {\b EAST}). 

However, they show very discrepant radial velocities. With {\sl Chandra},
{\b SOUTH} shows a best-fit redshift of (3.71$^{+0.24}_{-0.60}$),$\times$ 10$^{-2}$ consistent with the overall redshift 
determined optically (0.039$\pm$0.0003, Mohr et al. 1996\footnote{including all galaxy sub-populations 
discussed in Mohr et al. 1996}). The {\b EAST} region shows a much lower best-fit redshift of 
$\la$0.016 (the lower limits are not well constrained and are consistent with 0), implying a 
velocity difference of $>$ 3900 km~s$^{-1}$ at the 90\% confidence level. The velocity difference is consistent 
and better constrained than those obtained for similar regions with the {\sl ASCA} spectrometers.


{\sl XMM} MOSs analysis of the same regions show similar
velocity gradient. With MOS 2 the upper limit of the redshift values is not well constrained 
(there is a secondary $\chi^2$ minimum for the best-fit redshift at $\sim 0.035$). Since the overall results are 
very consistent between the two MOSs, we fitted MOS 1 \& 2 spectra simultaneously to improve statistics. 
The results of the simultaneous fittings
are also displayed in Table 1. The best fit redshift difference between these two regions is found to 
be $>$ 4000 km~s$^{-1}$ at the 90\% confidence level.

We can assess the statistical uncertainties of the velocity differences between 
these two region using the F-test, i.e.,
fitting the spectra of the two regions simultaneously with the redshifts locked together and comparing the 
resulting $\chi^2$ to that of simultaneous fittings where the redshifts are allowed to vary independently. 
The F-test indicates that the velocity differences in these two regions is significant at
the 99.8\%, 97.6\% confidence level for {\sl Chandra} ACIS-S3, and MOS 1 \& 2, respectively. 
The error-weighted average velocity difference from all three
detectors is (5.9$\pm$1.6)$\times$10$^3$km~s$^{-1}$ (the errors are 1-$\sigma$).

\subsection{
 Inclusion of Gain
                                 }
                                 
The significance of the velocity gradient described above only includes statistical uncertainties. The major source of 
uncertainty in velocity measurements with current spectrometers is the temporal and spatial 
variations of the instrumental gain. 
As in DB06, we can estimate the effects of residual gain fluctuations through Monte Carlo simulations. 
Given the relatively early date of the observations, 
we used the study of the gain variations
in the first 20 rows of {\sl Chandra} ACIS-S3 by Grant (2001) and assume that they also represent 
the expected variation for the MOSs as well. For a discussion on the gain stability 
in the {\sl XMM} detectors see Andersson \& Madejski (2004).

In order to assess the impact that random gain fluctuations would have on our results we simulated 500 spectra for
{\sl Chandra}, MOS 1 and MOS 2 using the XSPEC tool FAKEIT. The simulated spectra had the same input values as those
obtained through spectral fittings of the real data in regions {\b SOUTH} and {\b EAST} 
for N$_{H}$, temperature, oxygen, neon, magnesium, silicon, sulfur, argon, calcium, iron, nickel, normalization 
and were set at 
some intermediary redshift (z=0.029). The background and responses corresponded to that of the real data. Poisson
errors were included. The simulated spectra was then used to estimate the probability 
that a velocity difference similar or greater than
that observed in the real data in ACIS-S3, MOS 1 and MOS 2 could be generated by chance and how this
probability depended on the 
magnitude of gain fluctuations. The results are shown in Figure 4a, where we plot the probability that c(z$_{\b SOUTH}$ --
z$_{\b EAST}$) $>$ $\Delta$V as a function of the 1-$\sigma$ variation of the gain assumed 
300 km~s$^{-1}$ for individual velocity measurements, (Grant 2001)\footnote {There is evidence that both spatial and temporal
variations can be larger at later times (DB06).}.  We can see from Figure 4a 
that the significance of the velocity gradient is  $>$99\% assuming a 3-$\sigma$ gain variation. 


\subsection{
 Temporal and Spatial Gain Stability
                                 }
                                 
The two {\sl XMM} pointings from which the extraction regions were analyzed were taken with a separation of four days. 
We checked for possible anomalous gain variations that might have occurred between the two off-center 
observations by using 
a large elliptical region surrounding the cluster's center discussed in section 6.2 (seen in Figure 2a as 
the outer dashed lines). We fitted an absorbed APEC model and checked for redshift differences between different epochs in 
 MOS 1 \& 2 data individually. The best fit redshifts in the two epochs for MOS 1 are (3.98$\pm$0.39) 
 $\times$ 10$^{-2}$ and (3.61$\pm$0.26) $\times$ 10$^{-2}$. For MOS 2 the corresponding values are (3.56$\pm$0.39) 
 $\times$ 10$^{-2}$ and (3.66$\pm$0.13) $\times$ 10$^{-2}$. There were no significant changes in best-fit global 
 redshift between the two observations and also between different detectors.

Given the random variation of instrumental gain with position and time in the CCDs, it is useful to check whether
some particular CCD region has been more affected than others. Similarly to DB06,
we split the cleaned final ACIS-S3
event file into 3 different epochs (with $\sim$9.7 ksec each) and performed the
same velocity mapping as that described previously, i.e.,  through 
an adaptive smoothing routine that keeps a fixed minimum number of counts
per region (5000 counts) maintaining the range of fitting errors more or less 
constant for different regions. 
We then determined the standard deviation of the best fit velocities for the same region over 
different time periods. We plot the results in Figure 4b, where regions of high scatter are brighter.
The color steps in Figure 4b represent the average 1$\sigma$ fitting
 errors of the individual regions used to construct the velocity map. From Figure 2d, we can see that the 
 regions of significant low and high velocities are located in the zones with minimum redshift scatter ($\sigma_{z}\sim$0.004).
 This suggests that the velocity gradient is not dominated by local temporal variations of the gain.
That was the only instrument with enough counts to perform this analysis, given the loss of photons 
to flares with the {\sl XMM} data.


 \section{Individual Lines and Abundance Ratios
                                        }
   
Elemental abundance ratios can be used to determine the enrichment history of the intracluster gas 
(e.g. Mushotzky et al. 1996; Loewenstein \& Mushotzky 1996) and can, 
potentially, be used to characterize the ICM and to trace the 
origin of the undisturbed gas during merging (e.g. Dupke \& White 2003). This is because the internal variation
of these ratios is not random, but show typically a central dominance of SN Ia ejecta
(Dupke \& White 2000a,b; Finoguenov et al. 2000; Allen et al. 2001) \footnote {
Here we use the term SN Type dominance to denote SN Type Fe mass fraction, not to be confused with the actual number of
SNe.}. 
Dupke \& White (2003) have used the 
``lack'' of a chemical discontinuity in some cold fronts to point out that 
the scenario that cold fronts are 
caused by the unmixed remnant core of an accreted subsystem (Markevitch et al. 2002) is not the unique 
way to make cold fronts.
Here we use abundance ratios to test the merging
scenario, i.e.,
looking for a discontinuity that separates two different media with different enrichment histories.

Given the low temperatures and poor photon statistics for both {\sl Chandra} and {\sl XMM} observations the
abundances of silicon and iron are the best defined and isolated lines in the X-ray spectra 
in our usable frequency range. 
The Si/Fe ratio spans a relatively wide 
range of values between SN Ia and II yields, even when taking into account the theoretical yield uncertainties  
of different SN models (Gibson et al. 1997; Dupke \& White 2001a,b). Using the same adaptive smoothing routine as described
above we mapped the Si/Fe ratio throughout the cluster region with ACIS-S3. The results are shown in Figure 5a. The 
cluster's core sits on a clear separation between two media, highly discrepant in SN Type dominance. The Fe mass fraction 
towards the 
W and NW is strongly dominated by SN Ia ejecta while the E side is SN II ejecta dominated. The transition 
from SN Ia to II dominance is nearly centered along the arrow shaped brightness edge. 

Based on the Si/Fe {\sl Chandra} map we selected three characteristic regions for a direct comparison of the 
chemical enrichment gradient measured with {\sl Chandra} \& {\sl XMM}. These regions are circular and are denoted by 
CW (circle west),
C0 (circle center), CE (circle east) in Figure 1b. Individual silicon and iron abundances are shown in Table 2 and 
their ratios
derived from different instruments are plotted in Figure 5b. In Figure 5b we also show the theoretical limits for 
100\% SN II Fe mass fraction (top horizontal line) and 100\% SN Ia Fe mass fraction
for four theoretical supernova explosion models that differ in their explosion
characteristics (Nomoto et al. 1997a, b). The error weighted average of the SN Ia Fe mass
fraction contribution for CW is found to be 100$^{+0.00}_{-0.09}$\% as opposed to 33$\pm$22\% found for the CE region.


                                        \section{
Discussion
                                         }

In this work we re-analyzed the {\sl Chandra} observation of Abell 576 and determined the spatial distribution 
of temperatures, individual elemental abundances and radial velocities of the ICM, using the full field of view
of the ACIS-S3 and also two new {\sl XMM} observations covering similar spatial scales. This allowed us to compare the 
results obtained with different instruments having different systematic uncertainties. 
The velocity distribution near the core of the cluster shows a strong velocity gradient, in very good agreement 
both in magnitude and direction with the velocity gradient found with both SISs onboard
{\sl ASCA}. The error weighted average (over ACIS-S3, MOS 1 \& MOS 2) maximum velocity 
difference is found to be (5.9$\pm$1.6)$\times$10$^3$km~s$^{-1}$.
The combined set of observations makes the significance of velocity detection $>$99.9\% confidence,
when standard (1$\sigma$) gain fluctuations are taken into account. 

We also found a strong chemical gradient in the intracluster gas of this cluster. The distribution of iron 
and silicon abundances is asymmetric in such a
way as to produce a clear separation of the Si/Fe ratio at the cluster's center. If converted to SN Type enrichment,
the results indicate that nearly 67\% of the Fe mass has been produced by SN II towards the E and that the Fe mass 
content in the ICM towards the 
W and NW direction has been fully produced by SN Ia ($<$ 9\% produced by SN II).
This chemical gradient is very asymmetric, not consistent with the radial chemical gradients found
in some other clusters (e.g. Dupke 1998; Dupke \& White 2000a,b 2003; Finoguenov et al. 2000; Allen et al. 2001; De Grandi et al. 2004
Baumgartner et al. 2005). The general characteristics of this cluster are consistent with a 
merging origin as proposed by KD04. However, the velocity gradient in A576 suggests a larger line-of-sight component for the 
merger axis. 

The distribution of galaxy velocities in the field of A576 do not show any clear 
spatial segregation (Rines et al. 2000). However, the distribution of galaxies 
(from the NED database\footnote{nedwww.ipac.caltech.edu/}) with redshift within r$_{200}$ 
shows at least two large concentrations between 0.03$<$z$<$0.07 (Figure 6a). 
The first one is centered at z$\sim$ 0.0387, which is the characteristic cluster redshift. 
Since the velocity gradient found with {\sl Chandra} \& {\sl XMM} is very high we consider 
also the second galaxy clump at z$\sim$0.065. We separate three galaxy groups based on redshift: a 
low z group (0.03$<$z$<$0.0387), a high z group (0.0387$<$z$<$0.05) and a very high z group 
(0.057$<$z$<$0.07).
We plot the galaxies for these three groups in Figures 6b,c.  It can be seen from Figure 6b that the 
distribution of the 97 low z galaxies (blue) seem more isotropic than
that of the 76 high z galaxies (red) , which seems to be more concentrated towards the SW of the cluster. The
distribution of the 24 very high z galaxies (magenta) is displaced even more to the SW. Figure 6c shows a 
blow-up of Figure 6b with the velocity centroids of the three redshift groups (shown by ``X''s with the corresponding 
group colors). The velocity centroids are 2$^\prime$.4 (2$^\prime$.2) away form the X-ray center for the 
low (high) z group. The X-ray center is also $\ga$1$^\prime$.3 from the line connecting the centroids of the 
two groups, This difference is significantly out of the error ellipsoid for the velocity centroid (assuming 
6$\times$10$^{-5}$ and 2$^{\prime\prime}$.5 errors for redshift \& position, respectively (NED)). 

It is very difficult to make a direct comparison between the velocity measurements obtained from galaxy
 velocities and X-ray measurements given the difference of spatial scales. In general, the optical results are
  not inconsistent with the X-ray measurements. However, the absolute values
  between the redshifts of the galaxy concentrations and those obtained from X-ray spectroscopy are 
  discrepant and the results can only be compatible if there is an overall gain correction upwards. We
  do not have an external source to calibrate global gain corrections but it is unlikely that the same correction 
  would affect all three different instruments in different epochs. On the other hand, the methodology used here is sensitive
  to gain dependence on
  frequency (e.g., Dupke \& Bregman 2001b) and this is likely the reason for this discrepancy given the low temperatures
  of teh cluster (the redshift fitting process is weighted by the FeL complex). Even though the {\it absolute} 
  redshift values may be inaccurate, the redshift {\it differences} should not be affected, since the same 
  methodology was applied to all regions/and observations.
 So, we will assume that a correction of $\delta$z$\sim$0.015--0.02 should be applied to all measured redshifts
 when comparing the data in X-ray and optical frequencies.
  
  The orientation of the low--high velocity regions is
 very similar to that found in X-ray velocity measurements (NE--SW).  We also show the centroid of 
 the joint high \& very high z group in yellow. 
   The centroid of this group coincides with the most significant high velocity region (Figure 2d). 
   The above mentioned results using galaxy velocities can also be interpreted as due to an unusual amount 
   of interlopers (e.g. Wojtak \& Loas 2007) and in this section we discuss two scenarios that can explain
   the observations, i.e., projection of a background cluster and post-core crossing line of sight merging. 
   

                                        \subsection{
Projection Scenario
                                        }

The results presented above can be at least partially interpreted as resulting from a scenario where 
A576 is, in reality, 
two clusters closely aligned in the line of sight. The two clusters could be gravitationally unbound 
or in a pre-merger stage, in which case 
the velocity gradient would be mostly attributed
 to the clusters' Hubble distances. 
In this scenario the cores of both clusters would have to be near aligned in order to escape easy 
identification of a secondary
peak in surface brightness. 

Optical studies of A576 show several peculiarities that can be interpreted either as consequences 
of a cluster-cluster merging or as due to projection effects. Rines et al. (2000 - hereafter R00) used the kinematics 
of the infall region (Diaferio and Geller 1997) of 
Abell 576 to calculate the mass distribution out to several Mpc. Their method does not need the 
equilibrium assumptions typically used in X-ray mass estimations and relies on the fact that 
the velocity field around clusters is determined by the local dynamics of the dark matter halo. 
The amplitude of the characteristic ``trumpet shaped'' caustics in their velocity $\times$ radius plot 
is related to the escape velocity around halos. 
From their analysis one can infer that this cluster is passing through a major disturbance for several reasons,
among them, (1) a ``finger''
in phase space with high velocities for radii $<$ 2.9 $h_{70}^{-1}$ Mpc (Figure 4 of R00; see also Rines et al. 2003 and
Rines \& Diaferio 2006), 
(2) an apparent deficit of galaxies in the NW of the cluster (Figure 6 of R00),
(3) a similar geometrical configuration of high-velocity ``background'' system 
(centered nearly 8200 km~s$^{-1}$ over the cluster's redshift)
to the geometrical configuration of the cluster (Figures 14 \& 6 of R00), (4) an inferred total mass 2.5 times higher than that 
found from X-ray analysis in the same spatial scale
(see also Mohr et al. 1996). 

In order to estimate the likelihood that the velocity gradient is due to projection effects we
looked at the distribution of galaxy clusters
from cosmological N-body Hubble volume simulations. For that we use the positions of clusters in 
a 3 Gpc cube at z$\approx$0 selected in the data generated in Evrard et al. (2002). The virtual 
clusters were generated in a flat $\Lambda$CDM model, with $\Omega_m$ and $\Omega_\Lambda$ of 0.3 and 0.7, respectively 
and $\sigma_8$=0.9. 
Clusters were found using an algorithm that identifies halos as spheres, centered on 
local density maxima, with radii defined by a mean interior isodensity condition 
(see Appendix A of Evrard et al. 2002 for details). 

We searched within 500000 mock clusters those that had a projected core separation within 180 h$_{70}^{-1}$ kpc, 
 corresponding to 3.5$^\prime$ at a redshift $\sim$0.04. To 
be conservative we searched for a radial distance separation within 2$\sigma$ above and below the 
average redshift difference value of (5.9$\pm$1.6)$\times$10$^3$~km~s$^{-1}$. The results showed 265 systems that satisfied 
this criteria indicating a probability of 5$\times$10$^{-4}$ to
find such systems in the nearby universe.

                                        \subsection{
Merging Scenario
                                         }
                                         
Local mergers are, however, much more frequent. The same above mentioned Monte Carlo strategy 
applied to angular scales equivalent to the virial radius of a 
4 keV cluster, i.e., r$_{200} \sim 0.85 \sqrt{kT_{keV}}~h_{70}^{-1}$ Mpc~=~1.7 $h_{70}^{-1}$ Mpc, 
finds 3.9$\times$10$^4$ in 5$\times$10$^5$ clusters, i.e., a probability of 0.078. This 
estimate includes pairs of all relative velocities, but a recent analysis of
subhalo--host halo velocity differences  found for ``bullet clusters'' type 
(1E 0657-56 -- Markevitch et al. 2002) halos in the Millennium Simulation (Hayashi \& White 2006)
 indicates that large velocity differences are not uncommon. They find that 40\% of all host halos 
would have 1 out of the 10 most massive sub-halos 
with a velocity as high as that of the ``bullet cluster''. From these studies, we roughly estimate 
that the likelihood of an ongoing merger with sufficiently high relative velocity is at 
the percent level, and thus a few examples in the local population of observed massive 
clusters should be expected. 

The distribution of gas temperature, iron abundance, abundance ratios suggest that the merging axis 
component on
the plane of the sky would follow a NW-SE direction. 
The best configuration that explains the magnitude of the velocity gradient is a scenario similar to that 
of the ``bullet'' cluster (1E0657-56), i.e., a violent merger of two colder clusters and a (initial) merger 
axis making $\sim$ 80$^{\circ}$ with 
the plane of the sky and a small ($\sim$ 10$^{\circ}$, see below) deviation with respect to the N--S
direction\footnote{The closest configuration
with the ``bullet'' cluster would be a $\sim 180^{\circ}$ flip over the Y axis of Fig. 2 from Markevitch et al. (2002) 
where the observer is viewing from the left}.

A major prediction of the merging scenario is the presence of a hot ($>$10 keV if we scale from 
1E0657-56) component correspondent to the bow shock layer on the line of sight. 
In order to test the consistency of this prediction with the current data we extracted spectra from a large elliptical 
region surrounding the cluster's center covering the outer ``temperature ring'' seen in Figure 2a as dashed lines (but 
also including the center). We compared two spectral models fitting simultaneously 
five data sets, {\sl XMM} MOS 1 \& 2 data from the two pointings and 
ACIS-S3 data. The first one (model 1) was a single temperature {\tt WABS APEC}. The second (model 2) was a  
a double temperature {\tt WABS (APEC + APEC)} corresponding to the cold and hot components. The cold component temperature
was fixed at 3.5 keV, the lowest temperature observed throughout the temperature map.  
The normalization of the hot component was fixed at a fraction, f$_{norm}$ 
of that of the cold component. The number of degrees of freedom in the two models is the same given the constrains imposed to the 
double temperature component. We varied f$_{norm}$ from 1\% to 99\% and recorded the best-fit parameters. The results
are shown in Figure 7a, where we plot the $\chi^2$ distribution as a function of f$_{norm}$. It can be seen that the lowest $\chi^2$
is achieved at $\sim$25\% with a corresponding high temperature of 11.8 keV $<$ T $<$ 21 keV at the 1$\sigma$ level. 
From Figure 7a we can see that model 2 spectral fittings with f$_{norm}\ga$~12\% 
is better than those using a single temperature component (model 1), which has a 
$\chi^2$ of 1407 and is shown in Figure 7a as a straight line with a best fit temperature of 4.1 keV. 

For comparison, we estimated the fractional contribution 
of the hot component using a recently archived 100 ksec Chandra exposure of 
1E0657-56 (Observation ID 5356). In Figure 7b we show the raw X-ray image and the rectangular region used to 
extract a surface brightness profile along the main direction of motion of the ``bullet''
to estimate the relative emission measure. The size of the rectangular region 
($\sim$25$^{\prime\prime}$) corresponds
to $\sim$3$^\prime$ region in A576. On Figure 7c we show 
the surface brightness profile along the slice. From right to left the first surface brightness 
enhancement before the ``spike'' associated with the cool ``bullet'' is that of the shock region. 
Then, we see the colder 
``bullet'' followed by extended peak of the disturbed core of the primary cluster. 
The last component is a hot tail. We separated the regions in three parts based on the temperature 
map in Markevitch et al. (2002). The distribution of photon counts
for these three components (again from right to left) is approximately
1000 counts (shock region), 14500 counts (the two cold cores) and 3000 counts (hot tail),
which would place the f$_{norm}$ (hot/cold) at $\sim$ 26\% assuming that 
the bow shock symmetrically covers the two cluster cores. This fraction can be directly compared
to that derived using spectral fittings up to the precision of a (weak) function of temperature f(T)
($number counts \propto Projected Area \times Surf Brightness \propto density^2 \times f(T) \times Projected Area 
\propto  \frac {normalization_{VAPEC}}{characteristic~size} \times f(T)$). It is beyond the 
scope of this paper to carry out detailed modeling of 1E0657-56.
Nevertheless, we point out that the overall agreement of f$_{norm}$ with what would be expected from 
``seeing'' the ``bullet'' cluster along the merging axis is very consistent with a A576 passing through 
a near line of sight collision.

With the available data we do not have enough photon statistics and energy coverage to 
disentangle the multiple temperature components in the line 
of sight, i.e., cold gas from the pre-shocked ICM, a relatively thin bow shock, 
the projected high density cold cores, and finally 
the post and pre shocked material at the largest depth. However, we can roughly estimate a few 
merger parameters with the data at hand. From simple geometrical principles for a line of sight 
merger started at a time ``-t$_{shock}$'', the perturbation perpendicular to the surface of the Mach 
cone will propagate with the sound speed, so that 
the $cos \alpha=\frac{B}{c_{s}~t_{shock}}$, where $\alpha$ is half angle of the cone, 
c$_{s}$ is the sound speed given by 
$\sqrt{\frac{5kT_{ICM}}{3\mu m_{p}}} \approx 10^3~(\frac{T_{keV}}{3.7 keV})^{(\frac{1}{2})}$~km~s$^{-1}$, and
B the projected distance
from the merging axis to the point where the sonic perturbation is at a time $t_{shock}$.
Since the Mach number $M=\frac{1}{sin \alpha}$, the time when the shock front was effectively 
initiated is then $t_{shock}= \frac{B}{c_{s}\sqrt{1-M^{-2}}}$ or 
$t_{shock}\approx (0.08\pm0.015)~h_{70}^{-1}$ Gy ago, assuming B to be 
$B = 86 \pm 16~h_{70}^{-1}~kpc \sim 1^{\prime}.75\pm 0^{\prime}.5 $, where 2B$\sim$3$^{\prime}$.5 would be 
the projected distance between the two ``hot'' regions (NW \& E of the central region) in Figure 2a.
The distance traveled by the core along the line of sight 
during this time is $L\approx (0.45 \pm 0.15) h_{70}^{-1}$ Mpc for M=6$\pm$1.6, using the 
error-weighted average velocity derived from  {\sl Chandra} \& {\sl XMM} data.

The point in the past that the two merging clusters overcome the Hubble flow, with zero 
relative radial velocity (half the orbital period), 
can be given by 
$r_0 = (\frac{2G}{\pi^2})^{\frac{1}{3}} (M_c~t_{cross}^2)^{\frac{1}{3}}~\approx~5.5~(M_{c_{15}}~t_{cross_{Hub}}^2)^{\frac{1}{3}}$ Mpc,
where $M_{c_{15}}$ and $t_{cross_{Hub}}$ are the total mass normalized by $10^{15} M_{\odot}$ and $t_{cross_{Hub}}$ 
is the core crossing time normalized by a Hubble time (set to 1.37$\times10^{10}$yr). From conservation of energy and angular momentum the 
relative velocity of the sub-systems at a distance ``r'' from each other is given 
by (e.g. Ricker and Sarazin 2001) 

 \begin{eqnarray*}
v \sim \sqrt{2~G~M_c}~r^{-\frac{1}{2}}~(\frac{1-\frac{r}{r_0}}{1-(\frac{b}{r_0})^2})^{\frac{1}{2}}~\approx \\
4160\sqrt{M_{c_{15}}}~r_{0.5Mpc}^{-\frac{1}{2}}~(\frac{1-\frac{r}{r_0}}{1-(\frac{b}{r_0})^2})^{\frac{1}{2}} km~s^{-1},
 \end{eqnarray*}

where b is the impact parameter.
If we use the distance between the X-ray peak and the midpoint between the two ``hot'' regions ($=2B$) 
in Figure 2a as the 
impact parameter we obtain b=50$\pm$25~$h_{70}^{-1}$~kpc. Taking the total mass derived by Rines et al. (2000),
i.e., $M_c~=~(0.72 \pm 0.07) \times 10^{15} h_{70}^{-1} M_{\odot}$, the relative velocity at  $r~=~L$, when the merger shock is 
effectively initiated, is found to be (3.8 $\pm$ 0.63)$\times$10$^3$~km~s$^{-1}$.
This is in the lower end, but consistent, within the errors, with the observed 
velocity gradient, described in the previous paragraphs.

As pointed by Dupke \& Bregman (2002) and Sunyaev et al. (2003) ICM velocity detections can be 
corroborated by the use of the kinetic S-Z effect (Sunyaev \& Zel'dovich 1970, 1972, 1980). 
Intracluster gas bulk velocities as high as those detected in A576 should 
generate significantly different levels of Comptonization of the cosmic microwave 
background radiation (CMBR) towards different direction of the cluster (red-shifted and blue-shifted sides). 
 The total CMBR temperature variation towards the direction of a moving
cluster has a thermal and a kinetic component:

\begin {equation}
(\frac{\Delta T}{T})_{\nu}~ =~ [\frac{kT_{e}}{m_{e}c^{2}} (x \frac{e^{x} + 1}{e^{x} - 1} -4) - \frac{V_{r}(b)}{c}]~ \tau,
\end {equation} 

where $T_{e}$ \& $T$ are respectively the ICM and CMBR temperatures, V$_{r}$ is the radial velocity, $x=\frac{h\nu}{kT}$ and the other 
parameters have
their usual meanings (Sunyaev \& Zel'dovich 1970, 1972, 1980). If the gas 
number density $n(r)$ follows a king-like profile 
$n(r)= n_{0}(1 + (\frac{r}{r_{c}})^{2})^{- \frac{3}{2} \beta}$,
where ${r_{c}}$ and $n_{0}$ are respectively the core radius and the central density, 
the Thompson optical depth is given as a function of the projected radius ``$r_{proj}$'' by
$\tau(r_{proj}) = \sigma_{T} n_{0} r_{c} B(\frac{1}{2} , \frac{3}{2} \beta - \frac{1}{2})(1 +
(\frac{r_{proj}}{r_{c}})^{2})^{- \frac{3}{2} \beta + \frac{1}{2}}$,
where $B(p,~q)= \int_{0}^{\infty} x^{p-1} (1+x)^{p+q} dx$ is the Beta function of p, q.
Using $\beta$=0.64, $r_{c}$=240 $h_{50}^{-1}$ kpc, 
and $n_{0}$=2$\times$10$^{-3}$ cm$^{-3}$ (Mohr et al 1996), $\tau \sim 1.3 \times10^{-3}$ and
from equation (1) we get ($\frac{\Delta T}{T})_{217GHz}$~=~2.6$\times$10$^{-5}$, near the optimal 
frequency to observe the kinetic effect.
This effect could be detected with current (or in development) instruments, such as the 
{\sl BOLOCAM}\footnote{http://www.astro.caltech.edu/\~lgg/},
{\sl ACBAR} (Runyan et al. 2003), SuZIE (Holzapfel et al. 1997) or 
{\sl Planck}\footnote{http://www.rssd.esa.int/index.php?project=PLANCK\&page=index}.

The low photon statistics 
limits our ability to fully disentangle the 3-D physics of 
the merging event to make a close comparison to theoretical/numerical models. However, this work 
suggests that the temperature, abundance and velocity distributions in Abell 576 are consistent 
with a scenario where the cluster is passing through a line of sight merger similar to that in the ``bullet'' cluster.
If corroborated, this could provide a unique template to study supersonic line of sight cluster merger collisions.
This work also illustrates the power of elemental abundance gradient 
distribution in determining the evolutionary stage of clusters.

\acknowledgments 
The authors would like to thank Jimmy Irwin, Ed Lloyd-Davies, Maxim Markevitch, Chris Mullis, 
Kenneth Rines and Ming Sun for useful 
discussions and suggestions. We also thank the anonymous referee for useful suggestions. We acknowledge support
from NASA Grants NAG 5-3247, NNG05GQ11 \& GO5-6139X. This research made use of the HEASARC 
{\sl ASCA} database and NED.

                                
 \clearpage
                               \begin{figure}
                                \title{
Figure Captions
                                }
\caption{
(a) Raw {\sl Chandra} X-ray image of Abell 576. The X-ray contours shown here are used throughout the work.
North is up. The lowest contour is centered at RA=110.3762 deg, Dec=+55.7653 deg. The most external contour show the CCD borders 
and is limited by 110.5$<$RA$<$110.25 from left to right and 55.828$<$Dec$<$55.686 from top to bottom. The same contours 
are applied in Figures 2, 5b and 6a but with the scale slightly smaller. 
(b) Extraction regions used for spectral fittings for detailed analysis of radial velocities ({\b SOUTH} and {\b EAST}),
Si/Fe ratio (CW, C0, CE) analyzed in this work. We also indicate the regions found to have high radial velocities
(0$^{\circ}$--100$^{\circ}$) and low radial velocities (170$^{\circ}$--250$^{\circ}$) in a previous {\sl ASCA} analysis
(Dupke \& Bregman 2005a).
                                }
\caption{
Results from an adaptive smoothing algorithm with a minimum of 5000 counts per extraction 
circular region and fitted with an absorbed VAPEC spectral model. The gridding method used 
is a correlation method that calculates a new value for each cell in the regular matrix from 
the values of the points in the adjoining cells that are included within the 
search radius. With the minimum count constraints the matrix 
size was 50 $\times$ 50 cells. We also overlay the X-ray contours shown in Figure 1a on top of the 
contour plot). 
North is up. The lowest contour is centered at RA=110.3762 deg, Dec=+55.7653 deg. The units are pixels and 1 pixel=0.5 arcsec.
The arrow indicates 1 arcminute.
The parameters mapped are (a) Temperature (b) Redshift (c) Smoothed redshift error of each cell used in the adaptive binning 
(d) Deviation significance, i.e., redshift value found in
(b) minus the average for the whole CCD divided by the error of each measurement. The dashed ellipses shown 
in the Temperature plots indicate approximately the direction of the Mach cone in the scenario of 
near line of sight merger. The two stars near the center of the 
redshift map indicate the position of two bright E galaxies near the cluster's X-ray center, 
with relative line of sight velocity difference of 900 km/s (Smith et al. 2000). The average redshift error for 
each cell used in the adaptive binning code is 
is 0.01. The errors for the cells near the bottom left (SE) regions reach 0.02.
                                }
                                \caption{
(a)Best fit values for temperature, Fe abundance and redshift for the {\b SOUTH} and {\b EAST} regions shown in 
Figure 1b with different instruments. The left data point for instrument shows the value for {\b SOUTH} and the right 
data point the value for {\b EAST}. MOS 1\& 2 represent 
the results from simultaneous spectral fittings of the two MOS spectrometers.  
We also indicate the optically determined redshift for the cluster.
(b)TOP - Spectral fittings for regions SOUTH (white) and EAST (red) using Chandra ACIS-S3 data. 
BOTTOM  - A blow-up of the more prominent lines in the FeL and FeK complexes with the continuum subtracted.  
(c)Same as (b) but for the MOS 1 data.
(d)Same as (b) but for the MOS 2 data.
                                }
                                \caption{
(a)Probability of detecting a velocity difference greater than $\Delta$V for {\b SOUTH} and {\b EAST} 
regions. Solid line is without gain fluctuations. 
The other lines plots assume a 1$\sigma$, 2$\sigma$, 3$\sigma$, 4$\sigma$ and 5$\sigma$ 
gain fluctuation (500km s$^{-1}$ for individual velocity differences). Results are obtained from spectral
fittings of 500 simulated spectra for each region for  {\sl Chandra} and {\sl XMM}. (b) Smoothed map 
of the scatter (standard deviation) of the best fit redshifts over three time cuts (epochs)
each one having 9.5 ksec 
duration. Darker regions indicate lowest scatter and therefore higher gain stability. 
We also overlay the X-ray contours shown in Figure 1a on top of the 
contour plot). 
North is up. The lowest contour is centered at RA=110.3762 deg, Dec=+55.7653 deg. The units are pixels and 1 pixel=0.5 arcsec.
The arrow indicates 1 arcminute.
                                }
                               \caption{
(a)Results from an adaptive smoothing algorithm described in Figures 2 for the Si/Fe abundance
ratio found with {\sl Chandra} data. We also overlay the X-ray contours shown in Figure 1a on top of the 
contour plot). 
North is up.The lowest contour is centered at RA=110.3762 deg, Dec=+55.7653 deg. The units are pixels and 1 pixel=0.5 arcsec.
The arrow indicates 1 arcminute.
(b) Si/Fe abundance ratio measurements (by number normalized to solar) of Regions CW, C0 and CE using
{\sl Chandra} and {\sl XMM} MOS 1, 2 and 1\& combined.
We also shown the theoretical predictions for pure SN II enrichment (top horizontal line) and different models of pure SN Ia 
enrichment (standard W7 and Delayed Detonation models 1,2 \& 3  of Nomoto et al. (1997a, b)).
}
                               \caption{
(a) Histogram of galaxy velocities within a projected distance of 1 r$_200$ from the X-ray center. 
Data is from NASA/IPAC Extragalactic Database (nedwww.ipac.caltech.edu/). 
(b) Galaxy positions separated by redshift in the histogram shown in (a). Galaxies 
with redhifts 0.03$<$z$<$0.0387 are denoted by blue circles. Red circles denote galaxies with 
redshifts 0.0387$<$z$<$0.05 and magenta circles correspond to 0.057$<$z$<$0.07.
X-ray contours are also shown inthe center of the figure in white and the SOUTH and EAST 
boxy regions are shown in green. The large circle in black corresponds to $\sim$ 1 r$_200$.
(c) Blow-up of Figure 6b. Notation is the same as (b). It is also shown the velocity centroids for 
different redshift groups with ``X''. Blue corresponds to 0.03$<$z$<$0.0387, red to 0.0387$<$z$<$0.05, 
magenta to 0.057$<$z$<$0.07 and yellow to 0.0387$<$z$<$0.07.
}
                               \caption{
(a) $\chi^2$ variation of the best-fit double {\tt APEC} model to a large elliptical region encompassing the 
central regions of A576 as a function of the ratio of normalizations of the hot to cold components. Intermediate values of the 
best-fit high temperatures are shown for normalizations ratios of 10\%, 24\% (lowest $\chi^2$) \& 70\%. The 
temperature of the cold component was fixed at 3.5 keV.
The fit uses {\sl XMM} MOS 1 \& 2 data from the two off-center pointings and 
ACIS-S3 data simultaneously. The dotted lines show the results for a single APEC with a 
best-fit temperature of 4.1 keV, for comparison.
The number of degrees of freedom in the two models is the same given the constrains imposed to the 
double temperature component.
(b) ACIS-I image of 1E0657-56 from a deep (100 ksec) observation of the cluster. We also show the rectangular 
slice used to extract the surface brightness profile. North is up.
(c) Surface Brightness profile of the bullet cluster (1E0657-56) along the rectangular slice shown in Figure 7b.
The X-axis is shown in arcseconds and the Y-axis in arbitrary surface brightness units.
}
\end{figure}  

\begin{deluxetable}{lccccc}
\small
\tablewidth{0pt}
\tablecaption{Spectral Fittings for {\it SOUTH} \& {\it East} Regions\tablenotemark{a,b}}
\tablehead{
\colhead{Region/} &
\colhead{Temperature}  &
\colhead{Abund } &
\colhead{Redshift } &
\colhead{$\chi^{2}$/dof} & \\
\colhead{/Instrument} &
\colhead{(keV)} &
\colhead{(Solar)\tablenotemark{c}} &
\colhead{(10$^{-2}$)} &
\colhead{} &
}
\startdata
{\it SOUTH}/{\sl Chandra}& 3.75$^{+0.18}_{-0.18}$ & 0.47$^{+0.08}_{-0.08}$ & 3.71$^{+0.24}_{-0.60}$ & 578/398 \\
{\it SOUTH}/{\sl MOS 1}   & 4.05$^{+0.20}_{-0.28}$ & 0.60$^{+0.11}_{-0.16}$ & 3.72$^{+0.56}_{-0.52}$ & 728/429 \\
{\it SOUTH}/{\sl MOS 2}   & 3.70$^{+0.32}_{-0.32}$ & 0.71$^{+0.26}_{-0.16}$ & 4.76$^{+0.24}_{-0.66}$ & 728/429 \\
{\it SOUTH}/{\sl MOS 1\&2}& 3.95$^{+0.20}_{-0.20}$ & 0.62$^{+0.13}_{-0.13}$ & 4.18$^{+0.30}_{-0.30}$ & 728/429 \\
{\it EAST}/{\sl Chandra}  & 3.89$^{+0.25}_{-0.25}$ & 0.40$^{+0.09}_{-0.09}$ & 1.11$^{+0.48}_{-1.08}$ & 341/314 \\
{\it EAST}/{\sl MOS 1}     & 3.98$^{+0.22}_{-0.22}$ & 0.60$^{+0.13}_{-0.13}$ & 2.40$^{+0.56}_{-0.53}$ & 541/362 \\
{\it EAST}/{\sl MOS 2}     & 4.09$^{+0.20}_{-0.25}$ & 0.56$^{+0.07}_{-0.12}$ & 1.19$^{+2.73}_{-0.54}$ & 541/362 \\
{\it EAST}/{\sl MOS 1\&2}  & 4.03$^{+0.18}_{-0.18}$ & 0.58$^{+0.10}_{-0.10}$ & 1.87$^{+0.57}_{-0.20}$ & 541/362 \\
\enddata
\tablenotetext{a}{Errors are 1$\sigma$ confidence}
\tablenotetext{b}{Full energy range (0.5 keV--9.5 keV)}
\tablenotetext{c}{Photospheric}
\end{deluxetable}

\begin{deluxetable}{lcccc}
\small
\tablewidth{0pt}
\tablecaption{Individual Elemental Abundances \tablenotemark{a} }
\tablehead{
\colhead{Region/} &
\colhead{Silicon} &
\colhead{Iron}  &
\colhead{Si/Fe} & \\
\colhead{/Instrument} &
\colhead{(solar)} &
\colhead{(solar)} &
\colhead{} &
}
\startdata
{\it CW}/{\sl Chandra}& 0.26$\pm$0.26 & 0.80$\pm$0.12 & 0.33$\pm$0.33 \\ 
{\it CW}/{\sl MOS 1}   & 0.59$\pm$0.46 & 0.52$\pm$0.11 & 1.13$\pm$0.91 \\ 
{\it CW}/{\sl MOS 2}   & 0.32$\pm$0.32 & 0.61$\pm$0.09 & 0.53$\pm$0.53 \\ 
{\it CW}/{\sl MOS 1\&2} & 0.44$\pm$0.33 & 0.58$\pm$0.9 & 0.77$\pm$0.59 \\ 
{\it C0}/{\sl Chandra}& 0.89$\pm$0.23 & 0.73$\pm$0.07 & 1.23$\pm$0.33 \\ 
{\it C0}/{\sl MOS 1}   & 0.66$\pm$0.32 & 0.55$\pm$0.08 & 1.19$\pm$0.60 \\ 
{\it C0}/{\sl MOS 2}   & 1.07$\pm$0.32 & 0.71$\pm$0.08 & 1.50$\pm$0.48 \\ 
{\it C0}/{\sl MOS 1\&2}& 0.85$\pm$0.22 & 0.62$\pm$0.5 & 1.37$\pm$0.38 \\
{\it CE}/{\sl Chandra} & 1.38$\pm$0.38 & 0.43$\pm$0.12 & 3.20$\pm$1.25 \\
{\it CE}/{\sl MOS 1}   & 1.01$\pm$0.47 & 0.42$\pm$0.11 & 2.39$\pm$1.25 \\ 
{\it CE}/{\sl MOS 2}   & 0.90$\pm$0.45 & 0.44$\pm$0.10 & 2.05$\pm$1.13 \\ 
{\it CE}/{\sl MOS 1\&2}& 0.95$\pm$0.33 & 0.43$\pm$0.07 & 2.22$\pm$0.85 \\
\enddata
\tablenotetext{a}{Errors are 1$\sigma$ confidence}
\end{deluxetable}

                               \end{document}